\documentclass[]{spie}  
\usepackage{amsmath,amsfonts,amssymb}
\usepackage{graphicx}
\usepackage[colorlinks=true, allcolors=blue]{hyperref}
\graphicspath{{figures/}}

\providecommand{\minimize}{\mathop{\rm minimize}}

\providecommand{\subjectto}{\mathop{\rm subject\;to}}
\newcommand{\norm}[1]{\left\|{#1}\right\|} 
\newcommand{\lone}[1]{\norm{#1}_1} 
\newcommand{\ltwo}[1]{\norm{#1}_2} 
\newcommand{\lfro}[1]{\left\|{#1}\right\|_{\rm F}} 

\providecommand{\argmin}{\mathop{\rm argmin}}

\newcommand\bbR{\mathbb{R}}
\newcommand\eps{\varepsilon}

\newcommand\tr{\text{ tr}}

\title{A latent factor approach for prediction from multiple assays}

\author[a]{J. Kenneth Tay}
\author[b]{Robert Tibshirani}
\affil[a]{Department of Statistics, Stanford University, Stanford, CA 94305, USA}
\affil[b]{Department of Biomedical Data Sciences, and Department of Statistics, Stanford University, Stanford, CA 94305, USA}

\authorinfo{Further author information: (Send correspondence to J. Kenneth Tay)\\J. Kenneth Tay: E-mail: kjytay@stanford.edu}

\begin{document} 
\maketitle

\begin{abstract}
In many domains such as healthcare or finance,  data often come in different assays or  measurement modalities, with features in each assay having a common theme. Simply concatenating these assays together and performing prediction can be effective but ignores this structure. In this setting, we propose a model which contains latent factors specific to each assay, as well as a common latent factor across assays. We frame our model-fitting procedure, which we call the ``Sparse Factor Method'' (SFM), as an optimization problem and present an iterative algorithm to solve it.
\end{abstract}


\section{INTRODUCTION}\label{sec:intro}
Since the turn of the century, we have seen a rapid succession of technological advances which have enabled us to collect and store vast amounts of data. This data deluge promises to give us better understanding of the phenomena of interest. However such data can pose challenges for classical model-fitting algorithms. We need new algorithms which overcome these issues, and which exploit certain properties which this data might have.

For concreteness, let us suppose that we are in the biomedical context and that we would like to model a real-valued phenotype of interest, e.g. cholesterol level, which we denote by $y$. We collect various types of data on a number of patients, such as physical measurements, gene expression data, and survey results on patient lifestyle, which we hope are predictive of cholesterol level. Let us denote the data matrix for data source $k$ by $X_k$. If we have $n$ patients with $p_k$ variables from the $k$th data source, then $X_k \in \bbR^{n \times p_k}$.

The first issue we might face is the sheer number of features that we could use in our model. Assume for the moment that we are working with the column-wise concatenation of the $X_k$'s, denoted by $X$, which has a total of $p = p_1 + \dots + p_K$ features. The fact that we often have $p \gg n$ raises two problems. First, we might not be able to run our algorithm in the first place. As an example, ordinary linear regression and logistic regression are unidentifiable in this setting and will not give reasonable estimates. Second, we run into the issue of interpretability. It is not helpful to say that all $p$ features affect the response (``everything affects everything"). It would be preferable for a method to yield a model where the response $y$ depends on only a small subset of the $p$ features. This gives the researcher greater clarity on which features truly matter, as well as a specific shortlist of features to understand more deeply.

Statisticians have developed a host of methods that  yield a  \textit{sparse} fitted model, i.e. the response  is explained by a small subset of the features. The first general strategy is \textit{regularization}, that is, to impose an appropriate penalty on the coefficients in the model so that many of them are shrunk to zero.  This is the approach  taken by the LASSO \cite{Tibshirani1996}: $y$ is assumed to be a linear combination of the features, and an $\ell_1$ penalty is placed on the linear regression coefficients to induce sparsity. Other commonly used algorithms which promote sparsity include the elastic net \cite{Zou2005} and the Dantzig selector \cite{Candes2007}. A second general strategy is to add a screening step to remove features which are unlikely to have any relationship with the response before fitting the model. For example, the supervised principal components procedure \cite{Bair2006} removes features whose univariate standard regression coefficient is below a threshold.

The second issue we might face is how to pool information across the data sources. A general strategy we could adopt is to concatenate the $X_k$'s column-wise to get one large data matrix, $X$, and run our supervised learning method of choice with $X$. While this is a valid approach, it does not make use of any prior information we might have on the relevance of the data sources. Ideally, we would use an algorithm which exploits the group structure inherent in the data; at the least, it should respect this structure.

One approach is to introduce regularization penalties which involve the group structure. For example, the group LASSO \cite{Yuan2006} is an extension of the LASSO which, instead of promoting sparsity of coefficients for individual features, promotes sparsity of coefficients for groups of features. The sparse-group LASSO \cite{Simon2013} gives sparsity of groups as well as of features within the groups. However, these approaches might be too drastic for our application: thinking of the assays as groups, these methods will eliminate entire assays from the model.

Another approach we could take is model averaging. For each assay $X_k$, we fit a model, $\hat{f}_k$, for the response $y$ based on $X_k$. On new data $X_1^*, \dots, X_K^*$, we can predict the response via a weighted sum $\sum_{k=1}^K w_k \hat{f}_k(X_k^*)$, with the weights chosen via a suitable method (e.g. stacked generalization \cite{Wolpert1992}, \cite{Breiman1996}). We note that these methods do not allow the initial model-fitting to borrow strength across assays; this is a weakness if there is some underlying signal which is common across them.

In this paper, we present a model-fitting algorithm, called the ``Sparse Factor Method'' (SFM), which utilizes the group structure of the features and which results in sparsity of features. The method assumes that the data is generated from a latent factor model, and seeks to discover it via an optimization problem. In latent factor models, we assume the existence of unobserved variables, $U_1, \dots, U_r$, which are responsible for the signal present in both the response $y$ and the data $X$. Instead of seeking to explain the signal in $y$ using the features in $X$, we try to estimate the $U_k$'s and use them to predict $y$.

In Section \ref{sec:single}, we present the latent factor model for a single assay, our approach for fitting the model and simulation results. While this is not the primary setting of interest, the ideas and intuition behind the method are clearer in this simpler setting. The method extends naturally to the multiple-assay setting, which we describe, along with simulations, in Section \ref{sec:multiple}. Section \ref{sec:extension} generalizes the method to more complex latent factor models.

\section{DESCRIPTION OF METHOD FOR A SINGLE DATASET}\label{sec:single}

\subsection{Model description}
Denote the response of interest by $y \in \bbR^n$, and let $X \in \bbR^{n \times p}$ denote the data matrix of $p$ features from $n$ observations. While we have access to a large number of features (typically $p \gg n$), we typically expect only a small subset to have any association with the response. Let $P \subseteq \{ 1, \dots, p\}$ be the set of indices for these non-null features. For simplicity, we assume that both $y$ and the columns of $X$ have been scaled to have mean $0$ and variance $1$.

In this model, we assume that there is some underlying factor, $U$, which drives the response as well as the non-null covariates in a linear fashion, i.e.
\begin{equation}\label{eqn:latentfactor1}
y = \beta_0 + \beta_1 U + \eps,
\end{equation}
and
\begin{equation}\label{eqn:latentfactor2}
X_j = \begin{cases} \alpha_{0j}+\alpha_{1j} U +\eps_j &\text{if } j \in P, \\ \eps_j &\text{otherwise.} \end{cases}
\end{equation}

The errors $\eps$ and $\eps_j$ are assumed to have mean $0$ and to be independent of all other random variables in the model.

\subsection{Model-fitting as an optimization problem}
A plausible approach for fitting the model above is to first estimate $U$ and $\alpha$ as the first principal component of $X$ and its associated loadings, since it represents the direction of greatest variance in the design matrix. We can then run a linear regression of $y$ against the estimate of $U$ to get an estimate for $\beta$. However, this approach suffers from two problems. First, the loading vector $\hat{\alpha}$ is not sparse in general, meaning that the first principal component is a linear combination of all $p$ input variables. Second, the first principal component is determined in an unsupervised manner, without any reference to the response $y$. A method that uses signal information in the response could lead to better prediction.

The first issue can be overcome by insisting that the first principal component have a sparse loading vector. A number of different criteria have been suggested for defining sparse principal components (e.g. \cite{Jolliffe2003}, \cite{Witten2009}). One popular criterion is that of Zou et al. \cite{Zou2006}, where the loading vector $v$ solves
\begin{equation}\label{eqn:spca}
\minimize_{\alpha, v} \quad \lfro{X - X v \alpha^T}^2 \qquad \subjectto \quad \ltwo{v}^2 \leq 1, \lone{v} \leq c,  \ltwo{\alpha}^2 = 1.
\end{equation}
Here, $\lone{\cdot}$ and $\ltwo{\cdot}$ represent the usual $L_1$ and $L_2$ norms, while $\lfro{\cdot}$ represents the matrix Frobenius norm, i.e. $\lfro{X} = \sqrt{\tr (XX^T)} = \sqrt{\sum_{i,j} X_{ij}^2}$. $c$ is a hyperparameter which determines the sparsity of the solution vectors, with smaller values of $c$ corresponding to sparser solutions. It can fixed by some predetermined criterion, or selected via a method such as cross-validation. We could then set our estimate of $U$ to be $\hat{U} = X\hat{v}$, and run linear regression of $y$ against $\hat{U}$. 

To utilize the signal inherent in $y$ to better estimate $U$, we propose modifying \eqref{eqn:spca} to include the residual sum of squares (RSS) from fitting a linear model of $y$ against $\hat{U}$:

\begin{equation}\label{eqn:obj} \minimize_{\alpha, \beta, v} \quad \ltwo{y - X v \beta}^2 + w \lfro{X - X v \alpha^T}^2 \qquad \subjectto \quad \ltwo{v}^2 \leq 1, \lone{v} \leq c, \ltwo{\alpha}^2 = 1.
\end{equation}

In the above, $w$ is a tuning parameter which represents the relative weight that we give the reconstruction error of $X$ to the RSS of the linear regression of $y$ on $\hat{U}$. As $w$ increases, we give increasingly less weight to the signal in the response, and if we let $w \rightarrow \infty$, we end up with the sparse principal components criterion \eqref{eqn:spca}. In practice, $w$ can be chosen by cross-validation.

For computational efficiency, we remove the $L_2$ constraint on $v$:
\begin{equation}\label{eqn:triconvex}
\minimize_{\alpha, \beta, v} \quad \ltwo{y - X v \beta}^2 + w \lfro{X - X v \alpha^T}^2 \qquad \subjectto \quad \lone{v} \leq c, \ltwo{\alpha}^2 = 1.
\end{equation}

\subsection{Solving the optimization problem}\label{sec:singlesolve}

Let $f$ denote the objective function in \eqref{eqn:triconvex}. The objective $f$ is not jointly convex in its arguments and the constraints on the optimization problem are not convex,. However the optimization problem can be solved in an iterative fashion:

\begin{enumerate}
\item Initialize $\alpha$, $\beta$ and $v$.
\item Iterate until convergence:
\begin{enumerate}
\item $\beta \leftarrow \argmin_\beta f(\alpha, \beta, v)$.
\item $\alpha \leftarrow \argmin_\alpha f(\alpha, \beta, v)$ subject to $\ltwo{\alpha}^2 = 1$.
\item $v \leftarrow \argmin_v f(\alpha, \beta, v)$ subject to $\lone{v} \leq c$.
\end{enumerate}
\end{enumerate}

Each of Steps 2(a), 2(b), 2(c) can be solved more explicitly. Given $\alpha$ and $v$, $f$ is a quadratic function in $\beta$, which can be minimized by differentiating the objective function and setting the derivative to be 0. This yields $\beta^* = \frac{(Xv)^T y}{\ltwo{Xv}^2}$.

With $\beta$ and $v$ fixed, the optimization problem for $\alpha$ is equivalent to
\begin{equation}
\minimize_\alpha \quad \lfro{X - (Xv) \alpha^T}^2 \qquad \subjectto \quad \ltwo{\alpha}^2 = 1.
\end{equation}

We recognize this as the reduced rank Procrustes rotation problem which has a solution based on the singular value decomposition (SVD) of $X^T (Xv)$ (see Appendix \ref{app:procrustes} for details). In particular, since $X^T(Xv) \in \bbR^{p \times 1}$, we have an explicit solution $\alpha^* = \frac{X^T Xv}{\ltwo{X^T Xv}}$.

Given $\alpha$ and $\beta$, the optimization problem for $v$ is
\begin{equation}\label{eqn:v-bound}
\minimize_{v} \quad -2v^T X^T (w X\alpha + \beta y) + (w + \beta^2) v^T X^T X v \qquad \subjectto \quad \lone{v} \leq c.
\end{equation}
Letting $u = \frac{w X\alpha + \beta y}{w + \beta^2}$, (\ref{eqn:v-bound}) is equivalent to
\begin{align*}
&\minimize_v \quad -2 (Xv)^T u + \ltwo{Xv}^2 &&\subjectto \quad \lone{v} \leq c, \\ 
\Leftrightarrow\quad &\minimize_v \quad \| u - Xv \|_2^2 &&\subjectto \quad \lone{v} \leq c.
\end{align*}
We recognize this as the objective function which the LASSO solves, with $u$ and $X$ as the response and design matrix respectively and the constraint written in bound form. There is a fast implementation for obtaining a path of the solutions for the Lagrangian form \cite{Friedman2010}; solving the bound form simply involves finding the largest value of the dual parameter such that the $L_1$-norm constraint is satisfied.

With these simplifications, we can rewrite the iterative algorithm for fitting the model:

\begin{enumerate}
\item Initialize $\alpha$, $\beta$ and $v$.
\item Iterate until convergence:
\begin{enumerate}
\item $\beta \leftarrow \dfrac{(Xv)^T y}{\ltwo{Xv}^2}$.
\item $\alpha \leftarrow \dfrac{X^T Xv}{\ltwo{X^T Xv}}$.
\item $v \leftarrow \argmin_v \ltwo{u - Xv}^2$ subject to $\lone{v} \leq c$, where $u = \dfrac{w X\alpha + \beta y}{w + \beta^2}$.
\end{enumerate}
\end{enumerate}

We initialize $v$ at the loadings for the first principal component of $X$, and then we initialize $\alpha$ and $\beta$ based on this value of $v$. This seems to give good convergence in practice.

\subsection{Simulation study for one assay}\label{sec:singlesim}
We demonstrate the method outlined in Section \ref{sec:singlesolve} on simulated data. As we are interested in getting both accurate predictions as well as a sparse model, we benchmark the method against algorithms which satisfy these two criteria. Specifically, we compare our method with LASSO regression \cite{Tibshirani1996}, elastic net regression \cite{Zou2005} with the $\alpha$ parameter set to $0.5$, and supervised principal components analysis (PCA), introduced in \cite{Bair2006}. In a nutshell, supervised PCA keeps only features which are sufficiently correlated with the response, then performs linear regression of the response on the first principal component of the reduced data matrix.

In the first simulation, we simulate the latent factor model as in \eqref{eqn:latentfactor1} and \eqref{eqn:latentfactor2}. We set $n = 100$ and $p = 200$, with only $20$ of the features being non-null. Results can be seen in Figure \ref{fig:1assay}. Details on the simulation are given in Appendix \ref{app:1assay}. In both the low and high signal-to-noise ratio (SNR) settings, our method is competitive with supervised PCA in terms of test MSE, and outperforms elastic net and LASSO regression. Our method also has higher true positive rate than elastic net and LASSO regression, though perhaps at the expense of a higher false positive rate.

   \begin{figure} [h]
   \begin{center}
   \begin{tabular}{cc} 
   \includegraphics[height=5cm]{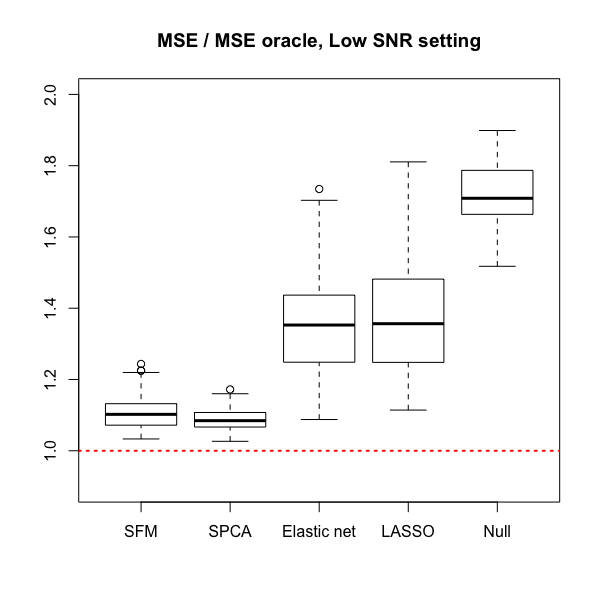} & \includegraphics[height=5cm]{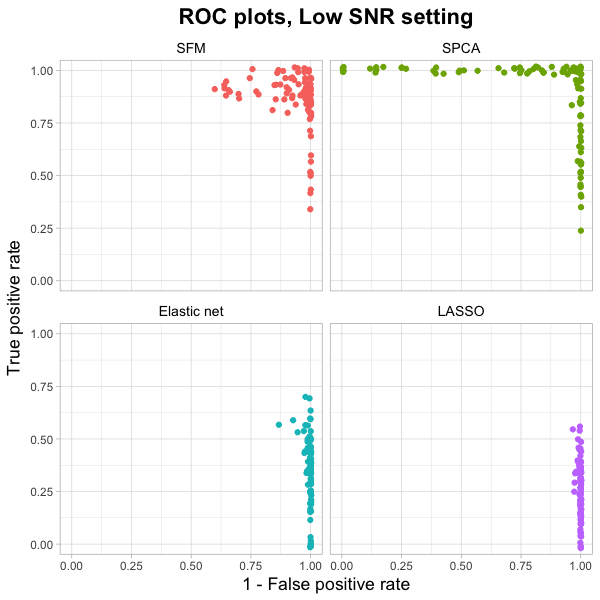} \\
   \includegraphics[height=5cm]{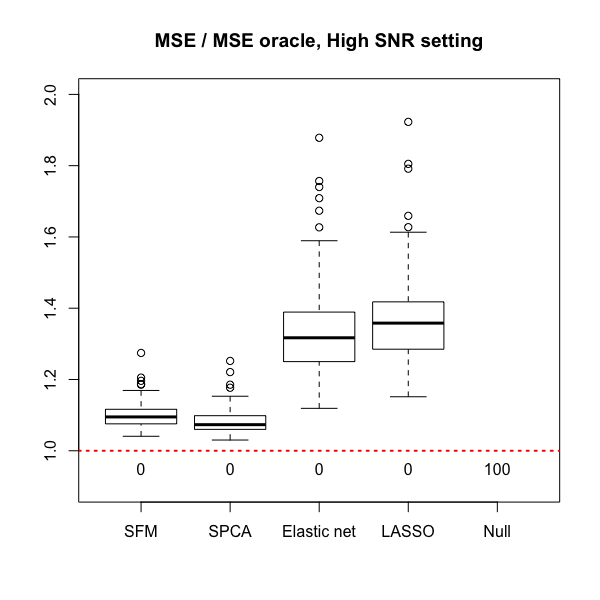} & \includegraphics[height=5cm]{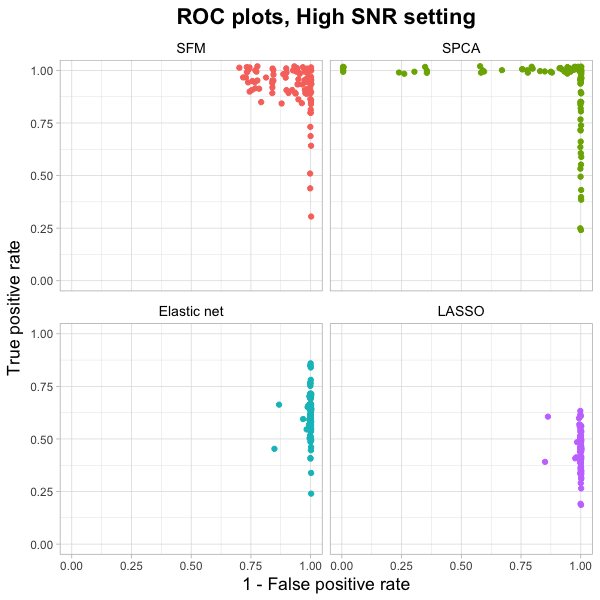}
   \end{tabular}
   \end{center}
   \caption
   { \label{fig:1assay} \textit{Data generated from the latent factor model \eqref{eqn:latentfactor1} and \eqref{eqn:latentfactor2}. On the left: comparing the test mean squared error (MSE) for the algorithms, normalized by the test MSE of the oracle which knows the underlying signal, for 100 simulated runs of data ($n = 100$, $p = 200$). On the right: comparing the ROC plots for the algorithms. In the bottom-left panel, the numbers represent the number of data points with values large enough that they do not appear on the figure. For our method, we set the hyperparameter $w = 0.2$.  In both the low and high SNR settings, SFM is competitive with supervised PCA in terms of test MSE, and performs better in terms of specificity.}}
   \end{figure}

\begin{figure} [h]
   \begin{center}
   \begin{tabular}{cc} 
   \includegraphics[height=5cm]{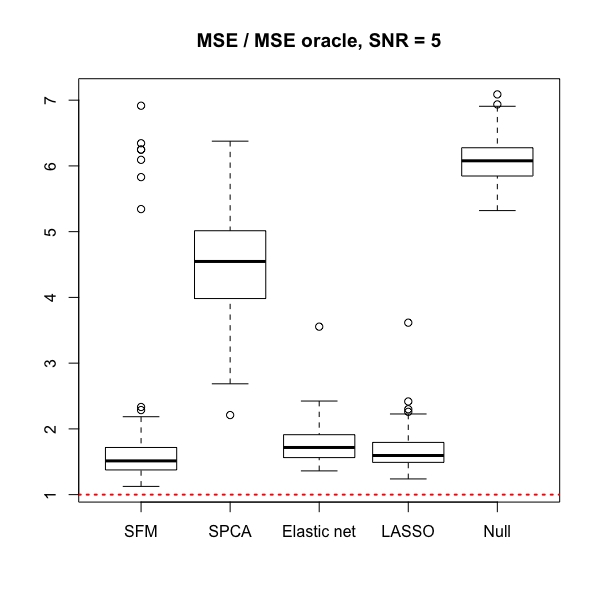} & \includegraphics[height=5cm]{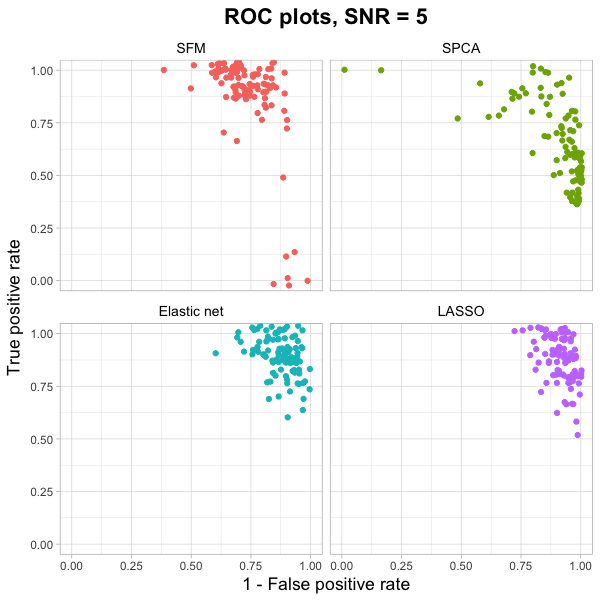}
   \end{tabular}
   \end{center}
   \caption
   { \label{fig:1assay_indep} \textit{$100$ independent features were generated, with the response being a linear combination of the first $10$ plus noise. On the left: comparing the test mean squared error (MSE) for the algorithms, normalized by the test MSE of the oracle which knows the underlying signal, for 100 simulated runs of data ($n = 100$). On the right: comparing the ROC plots for the algorithms. For our method, we set the hyperparameter $w = 0.2$. SFM remains competitive with elastic net and LASSO regression, while supervised PCA does not do well.}}
   \end{figure}

We consider a different setting in the next simulation. We set $n = 100$ and $p = 100$, and we assume that the features are independent, with the response being a linear combination of just the first $10$ features (details in Appendix \ref{app:1assay_indep}). Results can be seen in Figure \ref{fig:1assay_indep}. In this setting, supervised PCA performs poorly in terms of test MSE, while our method is still competitive with elastic net and LASSO regression. The true positive rate attained by our method is comparable to elastic net and LASSO regression, but has higher false positive rate as well.

\section{EXTENSION TO MULTIPLE ASSAYS}\label{sec:multiple}

\subsection{Model description}
The method developed in Section \ref{sec:single} extends naturally to the situation where we have multiple assays and latent factors which are both assay-specific and common across assays. Specifically, assume that we have $K$ assays $X_k \in \bbR^{n \times p_k}$, $k = 1, \dots, K$. For each $k$, let $P_k \subseteq \{ 1, \dots, p_k\}$ be the indices of the non-null features in assay $k$. Let $X_{k\ell}$ denote the $\ell$th column in the $k$th assay. We assume that the data is generated by the following latent factor model:
\begin{align}
y &= \beta_0 + \sum_{k = 1}^{K+1} \beta_k U_k + \eps, \text{ and } \label{eqn:multimodel1} \\
X_{k\ell} &= \begin{cases} \alpha_{0k\ell}+\alpha_{1k\ell} U_k + \alpha_{2k\ell} U_{K+1} +\eps_{k\ell} &\text{if } \ell \in P_k, \\ \eps_{k\ell} &\text{otherwise.} \label{eqn:multimodel2} \end{cases}
\end{align}

Note that $U_k$ is the latent factor specific to assay $X_k$, $U_{K+1}$ is a latent factor common across all the assays, and the response is a linear combination of all the latent factors in the model.

\subsection{Model-fitting as an optimization problem}

Let $p_{K+1} = p_1 + \dots + p_K$, and let $X_{K+1}$ be the column-wise concatenation of assays $X_1, \dots, X_K$. A natural extension of (\ref{eqn:obj}) for this set-up is
\begin{align}
\minimize_{\alpha, \beta, \gamma, v} \quad & \ltwo{y - \sum_{k=1}^{K+1} \beta_k X_k v_k}^2 + \sum_{k=1}^K w_k \lfro{X_k - X_k v_k \alpha_k^T - X_{K+1} v_{K+1} \gamma_k^T}^2 \label{eqn:multi-obj} \\ 
\subjectto \quad &\lone{v_k} \leq c_k, \quad k = 1, \dots, K+1, \nonumber \\ 
&\ltwo{\alpha_k}^2 = 1, \quad k = 1, \dots, K, \nonumber \\
&\ltwo{\gamma_k}^2 = 1, \quad k = 1, \dots, K. \nonumber
\end{align}

In the above, $\beta_1, \dots, \beta_{K+1} \in \bbR$, $v_k \in \bbR^{p_k}$, $\alpha_k \in \bbR^{p_k}$, and $\gamma_k \in \bbR^{p_k}$. The objective function is a linear combination of the RSS of the regression of $y$ on the estimated latent factors and the reconstruction errors for the $K$ assays, with the $w_k$'s indicating the relative weights of these errors.

\subsection{Solving the optimization problem}\label{sec:multisolve}
As in Section \ref{sec:singlesolve}, we can solve \eqref{eqn:multi-obj} in an iterated fashion. Let $f$ be the objective function in \eqref{eqn:multi-obj}, let $U_k = X_k v_k$ for $k = 1, \dots, K+1$, and let $U$ be the $n \times (K+1)$ matrix with the $U_k$'s as its columns. Fixing all other arguments, the minimization problem for $\beta$ is to minimize $\ltwo{y - U\beta}^2$. This is simply linear regression with response $y$ and data matrix $U$.

Fixing all other arguments, the minimization problem for $\alpha_1, \dots \alpha_K$ decouples into $K$ independent minimization problems:
\begin{equation} \minimize_{\alpha_k} \quad \lfro{ \left(X_k - U_{K+1}\gamma_k^T \right) - U_k \alpha_k^T}^2 \qquad \subjectto \ltwo{\alpha_k}^2 = 1.  \end{equation}

Each of these problems is the reduced rank Procrustes rotation with solution $\alpha_k^* = \frac{\left( X_k - U_{K+1}\gamma_k^T \right)^T U_k}{\ltwo{\left( X_k - U_{K+1}\gamma_k^T \right)^T U_k}}$. Similarly, fixing all other arguments, the minimization problem for $\gamma_1, \dots \gamma_K$ decouples into $K$ independent minimization problems of similar form:
\begin{equation} \minimize_{\gamma_k} \quad \lfro{ \left(X_k - U_k \alpha_k^T \right) - U_{K+1}\gamma_k^T}^2 \qquad \subjectto \ltwo{\gamma_k}^2 = 1.  \end{equation}

This is the reduced rank Procrustes rotation with solution $\gamma_k^* = \frac{\left( X_k - U_k\alpha_k^T \right)^T U_{K+1}}{\ltwo{\left( X_k - U_k\alpha_k^T \right)^T U_{K+1}}}$.

For $k = 1, \dots, K$, as a function of $v_k$,
\begin{align*}
f(v_k) &= \left( w_k + \beta_k^2 \right)\|U_k\|_2^2 + U_k^T \left( -2 \beta_k y + 2 \sum_{k' \neq k} \beta_k \beta_{k'} U_{k'} - 2 w_k X_k \alpha_k + 2 w_k U_{K+1} \alpha_k^T \gamma_k \right) \\ 
&= \left( w_k + \beta_k^2 \right)\|U_k\|_2^2 -2 U_k^T \left( \beta_k y - \sum_{k' \neq k} \beta_k \beta_{k'} U_{k'} + w_k X_k \alpha_k - w_k U_{K+1} \alpha_k^T \gamma_k \right).
\end{align*}

Completing the square, we see that minimizing $f(v_k)$ subject to $\lone{v_k} \leq c_k$ is equivalent to the LASSO problem in bound form with data matrix $X_k$ and response
\begin{equation}\label{eqn:v_k_response}
u = \dfrac{\beta_k y - \sum_{k' \neq k} \beta_k \beta_{k'} U_{k'} + w_k X_k \alpha_k - w_k U_{K+1} \alpha_k^T \gamma_k}{w_k + \beta_k^2}.
\end{equation}

As a function of $v_{K+1}$,
\begin{align*}
&f(v_{K+1}) = \left( \beta_{K+1}^2 + \sum_{k=1}^K w_k \right)\|U_{K+1}\|_2^2 \\ &\quad -2 U_{K+1}^T \left( \beta_{K+1}y - \sum_{k' \neq K+1} \beta_{K+1}\beta_{k'} U_{k'} + \sum_{k=1}^K w_k X_k \gamma_k - \sum_{k=1}^K w_k U_k \alpha_k^T \gamma_k \right).
\end{align*}

Hence, minimizing $f(v_{K+1})$ subject to $\lone{v_{K+1}} \leq c_{K+1}$ is equivalent the LASSO problem  with data matrix $X_{K+1}$ and response
\begin{equation}\label{eqn:v_K+1_response}
u = \dfrac{\beta_{K+1}y - \sum_{k' \neq K+1} \beta_{K+1}\beta_{k'} U_{k'} + \sum_{k=1}^K w_k X_k \gamma_k - \sum_{k=1}^K w_k U_k \alpha_k^T \gamma_k}{\beta_{K+1}^2 + \sum_{k=1}^K w_k }.
\end{equation}

In the summary, we can solve for \eqref{eqn:multi-obj} with the following iterative algorithm:
\begin{enumerate}
\item Initialize $\alpha_1, \dots, \alpha_K$, $\beta_1, \dots, \beta_{K+1}$, $\gamma_1, \dots, \gamma_K$, and $v_1, \dots, v_{K+1}$.
\item Iterate until convergence:
\begin{enumerate}
\item $\beta \leftarrow $ linear regression coefficients of $y$ on $U_1, \dots, U_{K+1}$.

\item $\alpha_k \leftarrow \frac{\left( X_k - U_{K+1}\gamma_k^T \right)^T U_k}{\ltwo{\left( X_k - U_{K+1}\gamma_k^T \right)^T U_k}}$ for $k = 1, \dots, K$.

\item $\gamma_k \leftarrow \frac{\left( X_k - U_k\alpha_k^T \right)^T U_{K+1}}{\ltwo{\left( X_k - U_k\alpha_k^T \right)^T U_{K+1}}}$ for $k = 1, \dots, K$.

\item For $k = 1, \dots, K$, $v_k \leftarrow \argmin_{v_k} \ltwo{u - X_k v_k}^2$ subject to $\lone{v_k} \leq c_k$, where $u$ is given by \eqref{eqn:v_k_response}.

\item $v_{K+1} \leftarrow \argmin_{v_{K+1}} \ltwo{u - X_{K+1}v_{K+1}}^2$ subject to $\lone{v_{K+1}} \leq c_{K+1}$, where $u$ is given by \eqref{eqn:v_K+1_response}.
\end{enumerate}
\end{enumerate}

\subsection{Simulation study for multiple assays}
We demonstrate the method outlined in Section \ref{sec:multisolve} on simulated data. As in Section \ref{sec:singlesim}, we compare the performance of our method against LASSO regression, elastic net regression with the $\alpha$ parameter set to $0.5$, and supervised principal components analysis. The three competitor algorithms are run on the concatenation of the three assays. For supervised PCA, we consider linear regression of the first 4 principal components of the reduced data matrix for better comparability to SFM.

In the first simulation, we generate data according to \eqref{eqn:multimodel1} and \eqref{eqn:multimodel2}. We have three  assays, each with $n = 100$ and $p = 100$, and only $10$ of the features in each assay being non-null. Results can be seen in Figure \ref{fig:multiassay}. Details on the simulation are given in Appendix \ref{app:multiassay}. SFM performs the best in terms of test MSE and true positive rate, though perhaps at the cost of a higher false positive rate.

   \begin{figure} [h]
   \begin{center}
   \begin{tabular}{cc} 
   \includegraphics[height=5cm]{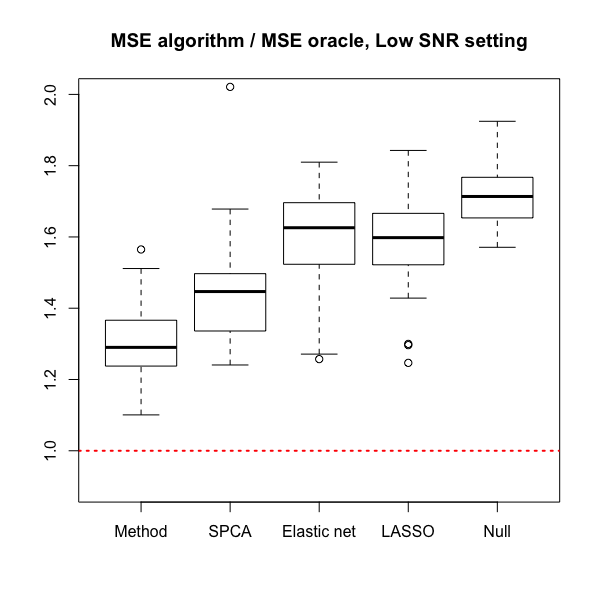} & \includegraphics[height=5cm]{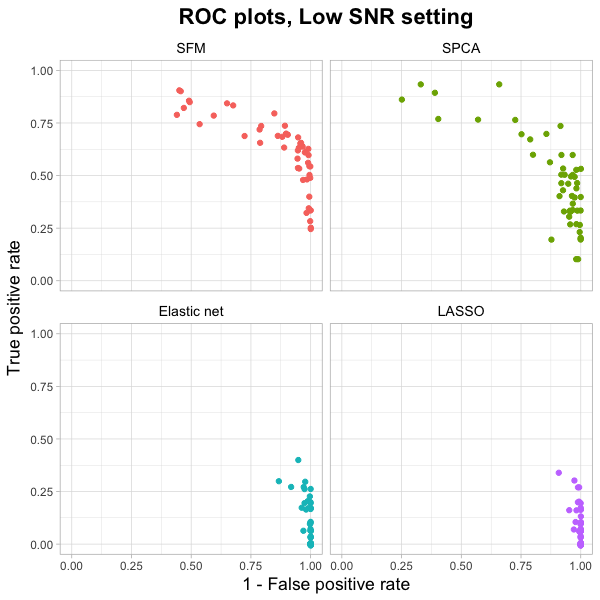} \\
   \includegraphics[height=5cm]{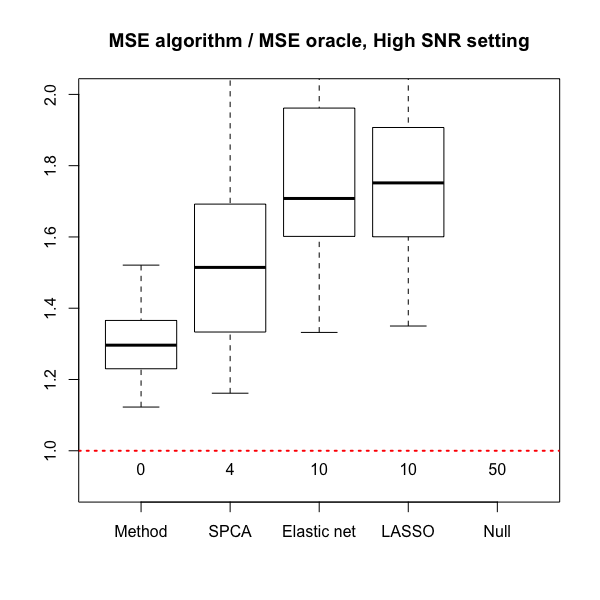}  &\includegraphics[height=5cm]{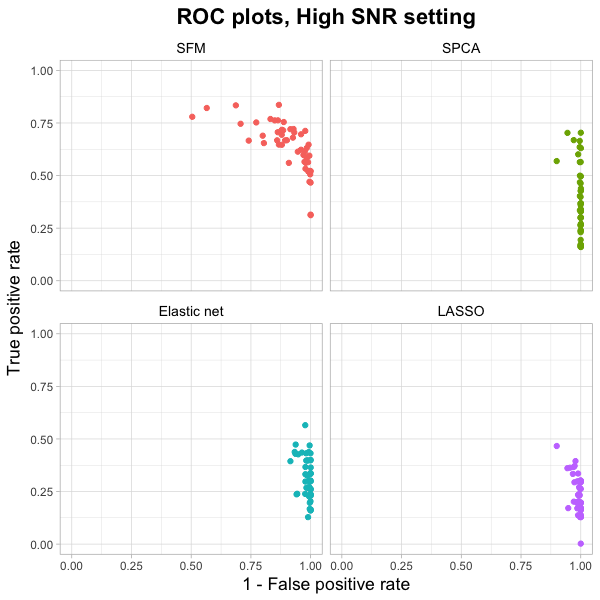}
   \end{tabular}
   \end{center}
   \caption
   { \label{fig:multiassay} \textit{Data generated from the latent factor model \eqref{eqn:multimodel1} and \eqref{eqn:multimodel2}. On the left: comparing the test mean squared error (MSE) for the algorithms, normalized by the test MSE of the oracle which knows the underlying signal, for 50 simulated runs of data. On the right: comparing the ROC plots for the algorithms. In the bottom-left panel, the numbers represent the number of data points with values large enough that they do not appear on the figure. For our method, we set the hyperparameter $w_k = 1$ for all $k$.  In both the low and high SNR settings, our method has a clear advantage in test MSE performance without sacrificing performance in sensitivity, although it seems to lose out in specificity.}}
   \end{figure}
   
\begin{figure} [h]
   \begin{center}
   \begin{tabular}{cc} 
   \includegraphics[height=5cm]{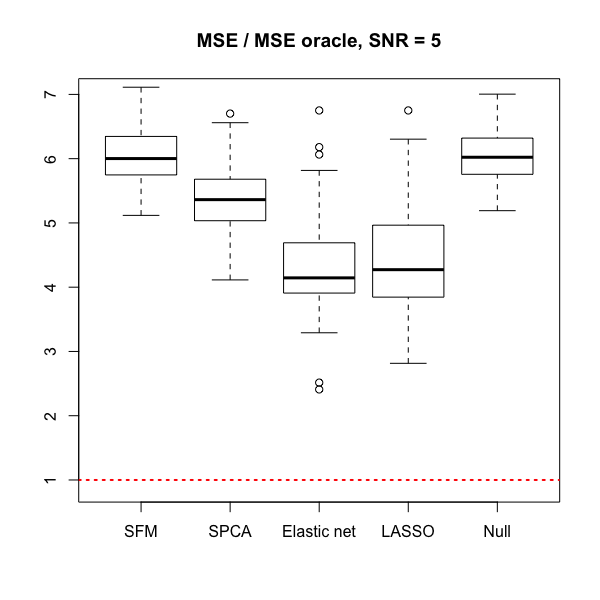} & \includegraphics[height=5cm]{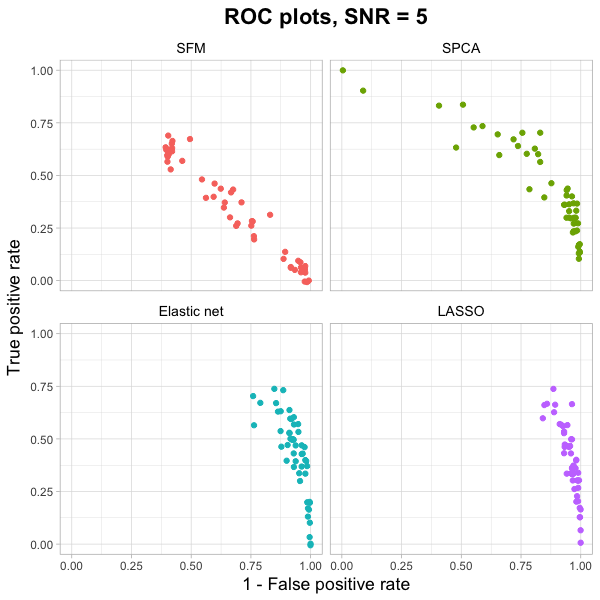}
   \end{tabular}
   \end{center}
   \caption
   { \label{fig:multiassay_indep} \textit{$100$ independent features were generated for each of three assays, with the response being a linear combination of the first $10$ in each assay plus noise. On the left: comparing the test mean squared error (MSE) for the algorithms, normalized by the test MSE of the oracle which knows the underlying signal, for 50 simulated runs of data ($n = 100$). On the right: comparing the ROC plots for the algorithms. For our method, we set the hyperparameter $w_k = 1$ for all $k$. Elastic net and LASSO regression perform best, as expected. Our method and supervised PCA not do well.}}
   \end{figure}

In the next simulation, we assume that we have three assays, each with $n = 100$ and $p = 100$. The features are all independent, with the response being a linear combination of the first $10$ features in each assay (details in Appendix \ref{app:multiassay_indep}). Results can be seen in Figure \ref{fig:multiassay_indep}. Elastic net and LASSO regression perform the best here; SFM and supervised PCA do not do well.

\section{GENERALIZATIONS TO DIFFERENT LATENT FACTOR MODELS}\label{sec:extension}
In Section \ref{sec:multiple}, we assumed that the underlying model had one latent factor common to all the assays and one latent factor specific to each of the assays. In this section, we consider extensions to more complex latent factor models.

\subsection{Multiple latent factors in one assay}
Assume for the moment that we only have one data assay $X$ available to us. Instead of assuming that just one latent factor underlying the data and the response, we can assume that there are $r > 1$ such factors:

\begin{equation}
y = \beta_0 + \beta_1 U_1 + \dots + \beta_r U_r + \eps,
\end{equation}
and
\begin{equation}
X_j = \begin{cases} \alpha_{0j}+\alpha_{1j} U_1 + \dots + \alpha_{rj} U_r +\eps_j &\text{if } j \in P, \\ \eps_j &\text{otherwise.} \end{cases}
\end{equation}

A natural extension of \eqref{eqn:triconvex} to fit this model is
\begin{align*}
\minimize_{A, \beta, V} \quad &\ltwo{y - X V \beta}^2 + w \lfro{X - X V A^T}^2 \\ 
\subjectto \quad & \lone{V_k} \leq c, \quad k = 1, \dots, r, \\
&A^T A = I,
\end{align*}
where $\beta \in \bbR^r$, $A, V \in \bbR^{p \times r}$.

Again, the optimization problem above yields to an iterative algorithm. For fixed $V$ and $A$, $\beta$ can be obtained by linear regression of $y$ on $XV$. For fixed $\beta$ and $V$, $A$ can be obtained by the reduced rank Procrustes rotation: if the SVD of $X^T XV$ is $RDS^T$, then the solution is $A^* = RS^T$. Since $X^T XV \in \bbR^{p \times r}$ with $r \ll p$, its SVD can be computed cheaply.

For each $k$, for fixed $\beta$, $A$ and $V_{k'}, k' \neq k$, by performing algebraic manipulations similar to that in Section \ref{sec:multisolve}, we can show that the optimization problem for $V_k$ is a LASSO problem.

\subsection{Multiple latent factors in multiple assays}
Next, let us assume that we have $K$ assays $X_1, \dots, X_K$, with $X_{K+1}$ denoting the column-wise concatenation of these assays. Let $X_k$ have $p_k$ features, and let $p_{K+1} = p_1 + \dots + p_K$. Let us assume that assay $k$ has $s_k$ latent factors which are specific to it, and that there are $s_{K+1}$ latent factors which are common to the $K$ assays. The natural optimization problem for this setting is
\begin{align*}
\minimize_{A, \beta, V} \quad &\ltwo{y - \sum_{k=1}^{K+1} X_k V_k \beta_k}^2 + \sum_{k=1}^K w_k \lfro{X - X_k V_k A_k^T - X_{K+1}V_{K+1} \Gamma_k^T}^2 \\ 
\subjectto \quad & \lone{V_{k,j}} \leq c_k, \quad k = 1, \dots, K+1, \quad j = 1, \dots, s_k, \\
&A_k^T A_k = I, \quad k = 1, \dots, K, \\
&\Gamma_k^T \Gamma_k = I, \quad k = 1, \dots, K,
\end{align*}
where $\beta_k \in \bbR^{s_k}$, $A_k, V_k \in \bbR^{p_k \times s_k}$, and $\Gamma_k \in \bbR^{p_k \times s_{K+1}}$.

This yields to an iterated algorithm that is very similar to that in Section \ref{sec:multisolve}. If we place a further restriction that $A_k^T \Gamma_k = 0$ for all $k$, we can solve for the $A_k$'s and $\Gamma_k$'s at the same time: the minimization problem becomes
\begin{align*}
\minimize_{A_k, V_k} \quad & \lfro{X - \begin{pmatrix} X_k V_k & X_{K+1} V_{K+1} \end{pmatrix} \begin{pmatrix} A_k & \Gamma_k \end{pmatrix}^T}^2 \\ 
\subjectto \quad & \begin{pmatrix} A_k & \Gamma_k \end{pmatrix}^T \begin{pmatrix} A_k & \Gamma_k \end{pmatrix} = I.
\end{align*}

This is again a reduced rank Procrustes rotation, but for a different matrix.

\section{Discussion and limitations}
Prediction and inference with multiple assays is a difficult problem. We wish to make use of the  group structure in the model-fitting algorithm, and we want a model which is sparse in its features within each assay but not in the assays. The Sparse Factor Method that we have introduced in this paper posits a latent factor model underlying the data, casts the model-fitting step as an optimization problem, and results in a model with the desired properties. SFM works by attempting to jointly minimize the reconstruction error of the assays after removing the estimated latent factors and the RSS of the linear regression of the response on the latent factors in the model.

There are challenges in this work  which would  benefit from further development. We list some here:
\begin{itemize}
\item The latent factor model implicit in the algorithm, while a reasonable starting point, may not be realistic in practice. For SFM to perform well, we would need to determine the number of latent factors to use. Even so, the response may not be related to the latent factors in a linear fashion.

\item There is the issue of how to choose appropriate values for the hyperparameters $w_1, \dots, w_K$, which represent the relative weights of the reconstruction errors. In theory, we could run cross validation to determine these values; in practice we often do not have sufficient data to do so.

\end{itemize}

We are in the process of distilling the model-fitting code into an R language package.

\appendix   

\section{Simulation details}\label{app:sim_details}

\subsection{Data generation for Figure \ref{fig:1assay}}\label{app:1assay}
\begin{enumerate}
\item $\beta \sim \frac{1}{2}\mathcal{N}(-3, 1) + \frac{1}{2}\mathcal{N}(3, 1)$.

\item For $j = 1, \dots, 20$, $\alpha_j \stackrel{iid}{\sim} \frac{1}{2}\mathcal{N}(-1.5, 1) + \frac{1}{2}\mathcal{N}(1.5, 1)$.

\item $U \sim MVN(0, I_{100})$.

\item \textbf{Generate response:} Compute $e_y^2$ using the relation $SNR_y = \dfrac{\text{Var} \left( \beta U \right)}{ \text{Var } \eps} = \dfrac{\beta^2}{e_y^2}$, then generate response $y = \beta U + \eps$, where $\eps \sim MVN (0, e_y^2 I_{100})$.

\item \textbf{Generate assay data:}
For $j = 1, \dots, 20$, compute $e_{xj}^2$ using the relation $SNR_x = \dfrac{\text{Var} \left( \alpha_j U \right)}{ \text{Var } \eps_j} = \dfrac{\alpha_j^2}{e_{xj}^2}$, then generate feature $X_j = \alpha_j U +\eps_j$, where $\eps_j \sim MVN (0, e_{xj}^2 I_{100})$.

For $j = 21, \dots, 200$, (i.e. $X_j$ is a null feature), $X_j \sim MVN(0, I_n)$.

\end{enumerate}

We set $SNR_x = SNR_y = 0.7$ in the low signal-to-noise ratio (SNR) setting, and $SNR_x = SNR_y = 2$ in the high SNR setting.

\subsection{Data generation for Figure \ref{fig:1assay_indep}}\label{app:1assay_indep}
\begin{enumerate}
\item For $j = 1, \dots, 10$, $\beta_j \stackrel{iid}{\sim} \frac{1}{2}\mathcal{N}(-2, 1) + \frac{1}{2}\mathcal{N}(2, 1)$. For $j = 11, \dots, 100$, $\beta_j = 0$.

\item \textbf{Generate assay data:}
For $j = 1, \dots, 100$, $X_j \stackrel{iid}{\sim} MVN(0, I_{100})$.

\item \textbf{Generate response:} Compute $e_y^2$ using the relation $SNR_y = \dfrac{\text{Var} \left( \beta_1 X_1 + \dots + \beta_{10}X_{10} \right)}{ \text{Var } \eps} = \dfrac{\beta_1^2 + \dots + \beta_{10}^2}{e_y^2}$, then generate response $y = \beta_1 X_1 + \cdots + \beta_{10}X_{10} + \eps$, where $\eps \sim MVN (0, e_y^2 I_{100})$.
\end{enumerate}

\subsection{Data generation for Figure \ref{fig:multiassay}}\label{app:multiassay}
\begin{enumerate}
\item For $k = 1, \dots, 4$, $\beta_k \stackrel{iid}{\sim} \frac{1}{2}\mathcal{N}(-3, 1) + \frac{1}{2}\mathcal{N}(3, 1)$.

\item For $k = 1, 2 , 3$, $j = 1, \dots, 20$, $\alpha_{jk}, \gamma_{jk} \stackrel{iid}{\sim} \frac{1}{2}\mathcal{N}(-1.5, 1) + \frac{1}{2}\mathcal{N}(1.5, 1)$.

\item For $k = 1, \dots, 4$, $U_k \stackrel{iid}{\sim} MVN(0, I_{100})$.

\item \textbf{Generate response:} Compute $e_y^2$ using the relation $SNR_y = \dfrac{\text{Var} \left( \beta_1 U_1 + \dots + \beta_{K+1}U_{K+1} \right)}{ \text{Var } \eps} = \dfrac{\beta_1^2 + \dots + \beta_{K+1}^2}{e_y^2}$, then generate response $y = \beta_1 U_1 + \cdots + \beta_{K+1}U_{K+1} + \eps$, where $\eps \sim MVN (0, e_y^2 I_{100})$.

\item \textbf{Generate assay data:}
For $k = 1,2,3$, $j = 1, \dots, 20$, compute $e_{xjk}^2$ using the relation $SNR_x = \dfrac{\text{Var} \left( \alpha_{jk} U_k + \gamma_{jk} U_{K+1} \right)}{ \text{Var } \eps_j} = \dfrac{\alpha_{jk}^2 + \gamma_{jk}^2}{e_{xjk}^2}$, then generate feature $X_{jk} = \alpha_{jk} U_k + \gamma_{jk}U_{K+1} +\eps_{jk}$, where $\eps_{jk} \sim MVN (0, e_{xjk}^2 I_{100})$.

For $k = 1,2,3$, $j = 21, \dots, 200$, (i.e. $X_j$ is a null feature), $X_{jk} \sim MVN(0, I_n)$.

\end{enumerate}

We set $SNR_x = SNR_y = 0.7$ in the low signal-to-noise ratio (SNR) setting, and $SNR_x = SNR_y = 2$ in the high SNR setting.

\subsection{Data generation for Figure \ref{fig:multiassay_indep}}\label{app:multiassay_indep}
\begin{enumerate}
\item For $k = 1, 2, 3$, $\ell = 1, \dots, 10$, $\beta_{k\ell} \stackrel{iid}{\sim} \frac{1}{2}\mathcal{N}(-2, 1) + \frac{1}{2}\mathcal{N}(2, 1)$. For $k = 1, 2, 3$, $\ell = 11, \dots, 100$, $\beta_{k\ell} = 0$.

\item \textbf{Generate assay data:}
For $k = 1, 2, 3$, $\ell = 1, \dots, 100$, $X_{k\ell} \stackrel{iid}{\sim} MVN(0, I_{100})$.

\item \textbf{Generate response:} Compute $e_y^2$ using the relation $SNR_y = \dfrac{\text{Var} \left( \sum_{k=1}^3 \sum_{\ell = 1}^{10} \beta_{k\ell} X_{k\ell} \right)}{ \text{Var } \eps} = \dfrac{\sum_{k=1}^3 \sum_{\ell = 1}^{10} \beta_{k\ell}^2}{e_y^2}$, then generate response $y = \sum_{k=1}^3 \sum_{\ell = 1}^{10} \beta_{k\ell}X_{k\ell} + \eps$, where $\eps \sim MVN (0, e_y^2 I_{100})$.
\end{enumerate}

\section{Reduced rank Procrustes rotation}\label{app:procrustes}

The reduced rank Procrustes rotation is stated as Theorem 4 of \cite{Zou2006}:
\begin{theorem}
Let $M_{n \times p}$ and $N_{n \times k}$ be two matrices. Consider the contrained minimization problem
\begin{equation}
\widehat{A} = \argmin_A \lfro{M - NA^T}^2 \qquad \subjectto \quad A^T A = I_{k \times k}.
\end{equation}
Suppose the SVD of $M^T N$ is $UDV^T$, then $\widehat{A} = UV^T$.
\end{theorem}

Since $M^T N \in \bbR^{p \times k}$, its SVD can be computed cheaply when $k$ is small.

\bibliography{report} 

\begin{thebibliography}{10}

\bibitem{Bair2006}
Eric Bair, Trevor Hastie, Debashis Paul, and Robert Tibshirani.
\newblock {Prediction by supervised principal components}.
\newblock {\em Journal of the American Statistical Association},
  101(473):119--137, 2006.

\bibitem{Breiman1996}
Leo Breiman.
\newblock {Stacked regressions}.
\newblock {\em Machine Learning}, 24(1):49--64, 1996.

\bibitem{Candes2007}
Emmanuel Candes and Terence Tao.
\newblock {The Dantzig selector: Statistical estimation when p is much larger
  than n}.
\newblock {\em Annals of Statistics}, 35(6):2313--2351, 2007.

\bibitem{Friedman2010}
Jerome Friedman, Trevor Hastie, and Robert Tibshirani.
\newblock {Regularization Paths for Generalized Linear Models via Coordinate
  Descent}.
\newblock {\em Journal of Statistical Software}, 33(1):1--24, 2010.

\bibitem{Jolliffe2003}
Ian~T. Jolliffe, Nickolay~T. Trendafilov, and Mudassir Uddin.
\newblock {A Modified Principal Component Technique Based on the LASSO}.
\newblock {\em Journal of Computational and Graphical Statistics},
  12(3):531--547, 2003.

\bibitem{Simon2013}
Noah Simon, Jerome Friedman, Trevor Hastie, and Robert Tibshirani.
\newblock {A Sparse-Group Lasso}.
\newblock {\em Journal of Computational and Graphical Statistics},
  22(2):231--245, 2013.

\bibitem{Tibshirani1996}
Robert Tibshirani.
\newblock {Regression Shrinkage and Selection via the Lasso}.
\newblock {\em Journal of the Royal Statistical Society. Series B
  (Methodological)}, 58(1):267--288, 1996.

\bibitem{Witten2009}
Daniela~M. Witten, Robert Tibshirani, and Trevor Hastie.
\newblock {A penalized matrix decomposition, with applications to sparse
  principal components and canonical correlation analysis}.
\newblock {\em Biostatistics}, 10(3):515--534, 2009.

\bibitem{Wolpert1992}
David~H Wolpert.
\newblock {Stacked Generalization}.
\newblock {\em Neural Networks}, 5(2):241--259, 1992.

\bibitem{Yuan2006}
Ming Yuan and Yi~Lin.
\newblock {Model Selection and Estimation in Regression with Grouped
  Variables}.
\newblock {\em Journal of the Royal Statistical Society. Series B (Statistical
  Methodology)}, 68(1):49--67, 2016.

\bibitem{Zou2005}
Hui Zou and Trevor Hastie.
\newblock {Addendum: Regularization and variable selection via the elastic
  net}.
\newblock {\em Journal of the Royal Statistical Society: Series B (Statistical
  Methodology)}, 67(2):301--320, 2005.

\bibitem{Zou2006}
Hui Zou, Trevor Hastie, and Robert Tibshirani.
\newblock {Sparse principal component analysis}.
\newblock {\em Journal of Computational and Graphical Statistics},
  15(2):265--286, 2006.

\end{thebibliography}
\bibliographystyle{plain}

\end{document}